\RequirePackage{fix-cm}
\documentclass[twocolumn]{svjour3}          
\usepackage{graphicx}
\usepackage{color}
\usepackage{amsmath}
\usepackage{gensymb}
\journalname{Granular Matter}
\begin{document}

\title{Stagnant Zone Formation in a 2D Bed of Circular and Elongated Grains under Penetration}



\author{Matt Harrington         \and
		Hongyi Xiao				\and
        Douglas J. Durian
}


\institute{M. Harrington $\cdot$ H. Xiao $\cdot$ D.J. Durian\at
              Department of Physics and Astronomy\\             
              University of Pennsylvania\\
              Philadelphia, PA 19104 USA\\
              \email{djdurian@physics.upenn.edu}           
}

\date{Draft: \today}

\maketitle

\begin{abstract}
When a large blunt object penetrates into a granular material, the force response exhibits an initial nonlinear relationship with depth that is widely attributed to the formation of a shear jammed stagnant zone. We present an experimental study of a model 2D granular system that preserves common aspects of slow granular penetration, such as a linear force law for large depths, while also allowing for direct and precise measurement of a stagnant zone. We also probe structural anisotropies throughout the bed and examine their characteristics inside and outside the stagnant zone. Finally, we simultaneously consider effects of concave elongated particle shapes (dimers) and find that while most results are not affected, the relationship between local structure and strain suggests a distinct fragility in the stagnant zone.
\keywords{Penetration \and Impact \and Shear jamming \and Particle shape}
\end{abstract}

\section{Introduction}
\label{intro} 

We are honored to contribute to this special issue dedicated to the memory of Bob Behringer. For several years, Bob studied impact and penetration into granular materials and other complex fluids~\cite{GengPRL2001,DanielsChaos2004,GengPRE2005,KondicPRE2012,ClarkPRL2012,ClarkEPL2013,ClarkPRE2014,ClarkPRL2015,ClarkPRE2016,BesterPRE2017,LimEPL2017,LimPRL2017,ZhengPRE2018}, as well as the dynamics and rigidity of packings of particles that are elongated~\cite{FarhadiPRL2014,FarhadiPRL2015,TangEPL2016} or interlocked~\cite{ZhaoGM2016}. Here we present new work lying at the interface between these two topics. 

One of the most ubiquitous and longest studied processes in the field of granular matter is the response of the grains to a penetrating object~\cite{AllenJAP1957,RuizSuarezRPP2013,HiroakiBook,DevarajARFM2017}. The impact process can dissipate the intruder's kinetic energy~\cite{UeharaPRL2003,AmbrosoPRE2005a,AmbrosoPRE2005,KatsuragiNatPhys2007,GoldmanPRE2008,ClarkPRL2012,NordstromPRL2014,ClarkPRL2015,BesterPRE2017} or be driven at constant speed~\cite{GengPRE2005,BrzinskiPRL2013}, and can even include an interstitial fluid that introduces cohesion~\cite{Thoroddsen2012,TedPRE2015} and aspects of shear thickening~\cite{LimPRL2017,BrownJaeger2014,WaitukaitisNature2012,HanNatCommun2016}. The implications of this research broadly reach into disparate fields and applications, from growth of plant roots in soils~\cite{Bingham2003,Kolb2012,Wendell2012,SchunterPRL2018}, to astrophysical impact cratering~\cite{Melosh1999,Walsh2003,Lohse2004,DanielsChaos2004,GuttlerIcarus2012,HousenIcarus2018}, to robotic systems interacting with complex media~\cite{LiPNAS2009,McInroeScience2016,AguilarNatPhys2016,Drotman2017}.  

A common and crucial objective of these studies is to model the resistive force that the grains exert on the intruder~\cite{KatsuragiNatPhys2007}. In general there is rate-dependent inertial drag plus a quasi-static force that increases with depth and originates from gravity-loaded frictional contacts~\cite{BrzinskiPRL2013}. The latter is of the form $d{\bf F}=-\alpha \mu (\rho g z){\rm d}{\bf A}$ where $\mu$ is an internal friction coefficient equal to the tangent of the repose angle, $\rho g z$ is the granular hydrostatic pressure at depth $z$, $d{\bf A}$ is an infinitesimal area element pointing normal to the projectile surface, and $\alpha$ is the proportionality constant, found to be $\alpha=35\pm 5$ \cite{BrzinskiPRL2013}.  The surprisingly large value of $\alpha$ may be due to the mobilization of friction along force chains or the stagnant zone (next) that protrudes ahead of the projectile.

Another critical aspect of granular penetration involves the development of a so-called stagnant zone. As an object moves through a granular bed, a locally jammed region tends to form directly in front of the object~\cite{AlbertPRL2000,StoneNature2004}. For a vertically penetrating object with a flat bottom, the stagnant zone manifests as a conical region of constrained grains that acts like a plow into the granular bed. This is thought to be an instance of shear jamming~\cite{BiNature2011}, only occurring for frictional grains. Indeed, regions of low strain resembling stagnant zones have been observed in quasi-2D experiments~\cite{AguilarNatPhys2016,MurthyPRE2012,ViswanathanGM2015}. Recently, both the local friction force framework and stagnant zone formation have been incorporated into a force model inspired by Archimedes' principle for buoyant fluids~\cite{KangNatComms2018}. Still, further experimental investigation that can capture both the global force response and stagnant zone formation is needed.

In this article, we describe a new study that examines a model 2D granular system under steady penetration, in an environment that seeks to preserve aspects of penetration into 3D beds, while also allowing for precise measurements of structure and motion. In Sec.~\ref{sec:methods}, we describe the materials and methods used in this study. We also describe the granular systems we use to vary particle shape, examining packings of particles that are either circularly symmetric or concave elongated. In Sec.~\ref{sec:results}, we summarize the primary findings of this study. Namely, we examine the force response in Sec.~\ref{sec:forces}, stagnant zone detection in Sec.~\ref{sec:SZ}, and local structure and its relationship to local strain in Sec.~\ref{sec:structure}. We close with a discussion of the results and prospects for future study in Sec.~\ref{sec:discussion}.  

\section{Methods \& Materials}
\label{sec:methods}

To begin, we describe the experimental apparatus and granular materials, as well as the basic analysis tools for force readings and particle tracking. The tools used here are derived from prior studies on uniaxial compression of granular pillars~\cite{CubukSchoenholzPRL2015,LiRieserPRE2015,RieserPRL2016,HarringtonPRE2018}.

\subsection{Penetration Apparatus}
\label{sec:apparatus}

The experimental apparatus consists of a table-top box, 27~in.\ x 23~in.\ (69~cm x 58~cm), with a translating stage that moves along rotating rods and a dry acrylic substrate. At the bottom of the box, a 3-sided rigid frame is placed on the substrate, its bottom in contact with a static bar. The inner width of the frame is 18~in.\ and the side bars have a length of 5~in.  The lateral dimensions are chosen to avoid boundary effects from the side bars~\cite{SeguinPRE2008,NelsonPRL2008}. Consideration of the bottom boundary is discussed in detail in Sec.~\ref{sec:SZ}. A granular material, described in full in Sec.~\ref{sec:grains}, is randomly and closely packed into the frame as a monolayer. Cable tie heads are fixed to the bottom of the frame in order to frustrate structural order and sliding motion along the bottom of the packing. 

\begin{figure}
\centering
\includegraphics[width=0.70\linewidth]{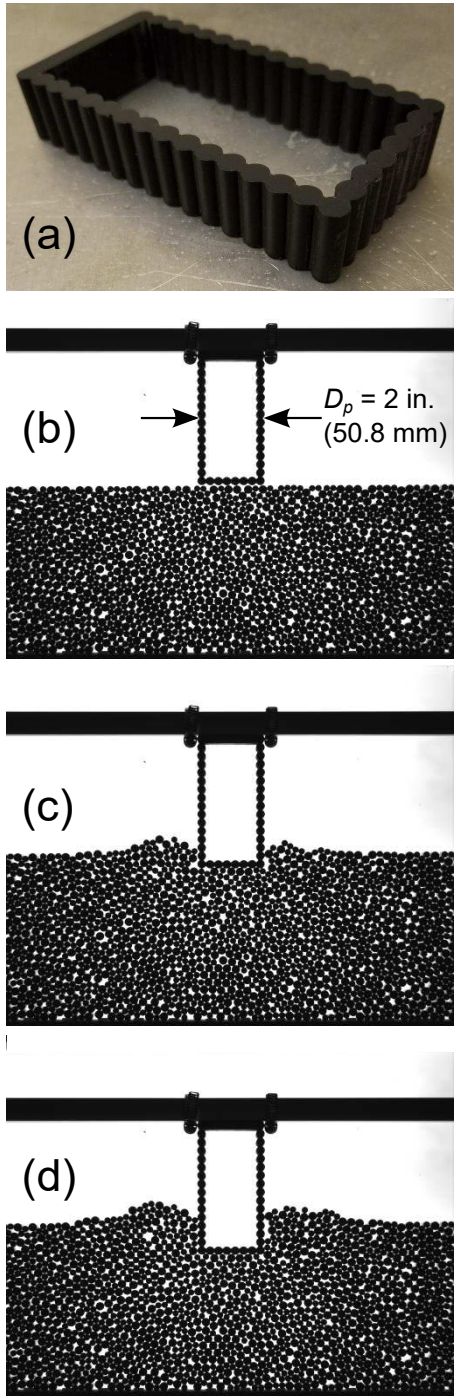}
\caption{(a) Photograph of our custom 3D printed intruder. The outer dimensions are 4~in.\ x 2~in.\ x 0.75~in.\ (102~mm x 51~mm x 19~mm). The intruder design includes an array of connected circular rods with diameter 0.25~in.\ (6.4~mm), matching that of large monomers. The circular bumps are tracked as if they are additional grains. The bottom row acts as an array of `frozen' grains that encourage the formation of a stagnant zone. The bumps along the sides do not affect the global force behavior. (b)-(d) Raw images captured during penetration into a packing of monomers at depths of 0, $2D$, and $4D$, respectively, where $D$ is the diameter of a large monomer.}
\label{fig:apparatus_intruder}
\end{figure}

The grains are driven by a penetrating object that slides along the substrate, in contact with a suspended horizontal bar carried on the translating stage. The intruding object, as shown in Fig.~\ref{fig:apparatus_intruder}(a), is 3D printed (Formlabs Form~2) using a photopolymer resin (Formlabs FLGPCL03). The shape of the object is designed to contain circular bumps that are 0.25~in.\ (6.4~mm) in diameter, the same diameter as large circular grains in the system, along three of the sides. The choice in this design affords ease in tracking the intruder, characterizing strain and structure in the vicinity of the intruder, and encouraging the formation of a stagnant zone as the bottom row acts as `frozen' grains. The bumps on the side walls do not affect global force behavior, as the contribution from frictional forces act normal to surface elements of an intruding object, rather than tangentially~\cite{BrzinskiPRL2013}. The bounding box dimensions of the intruder are 4~in.\ x 2~in.\ x 0.75~in.\ (102~mm x 51~mm x 19~mm). At the beginning of an experiment, after the grains are packed into the frame, the flat side of the intruder is placed on the substrate and in contact with the moving bar. The stage, carrying the bar that pushes the intruder, is set into motion toward the granular material with a speed of 0.0033~in./s (85~$\mu$m/s). 

While the apparatus is active, force readings and images are recorded. Two force sensors (Omega Engineering LCEB-5) are mounted onto the translating stage, reading the force that, effectively, the intruder feels. Meanwhile, a camera (JAI/Pulnix TM-4200CL) mounted above the experimental apparatus images the system from the top-down. Acquired images are 8-bit and 4.2 megapixel (2048 x 2048) with a resolution of 77~pixels/in.\ (3~pixels/mm). While the full depth of grains is imaged, the imaged width is roughly 13.25~in.\ (34~cm). Force readings and images are acquired simultaneously at a rate of 8~Hz. Typical raw images captured during an experiment are shown in Fig.~\ref{fig:apparatus_intruder}(b)-(d).

\subsection{Granular Materials}
\label{sec:grains}

The granular systems consist of dry bidisperse (1:1) acetal (Delrin) rods. Large rods have a diameter, $D$, of 0.25~in.\ (6.4~mm) and small rods have a diameter, $0.75D$, of 0.1875~in.\ (4.8~mm), while both species have a height of 0.75~in.\ (19~mm). The rods are placed on the substrate standing upright, creating a model two-dimensional (2D) granular system. Since the apparatus rests on a table-top, gravity points into the substrate so it does not play a role other than to set friction with the substrate. When driven by the intruding object, the rods slide along the substrate, with a friction coefficient of $0.23 \pm 0.01$~\cite{RieserThesis}. 

The rods are used to study different shapes of particles. As individual, discrete rods, we refer to them as monomers. We also study systems in which every rod is adhesive-bonded with a like-sized partner, which we refer to as dimers. This creates an elongated particle shape, which can profoundly affect penetration deformation in other disordered solids~\cite{ZhangACSNano2013}. These specific granular shapes have previously been studied in the context of uniaxial compression~\cite{HarringtonPRE2018}. While the 18~in.\ width is explicitly held fixed, the initial $\approx 5$~in.\ height of the granular system is varied between trials, and especially between the two shapes. To keep the physical dimensions of the granular system consistent, we also vary the total number of particles, $N = 2000$ for monomers, and $N = 1000$ for dimers.

\subsection{Force Analysis}
\label{sec:forceAnalysis}

As mentioned in Sec.~\ref{sec:apparatus}, we record the force the intruder feels while sliding into the granular system, at a rate of 8~Hz, using two force sensors mounted on the translating stage. Since the intruder begins an experiment well in front of the granular material, we subtract the average initial signal in the two sensors, which corresponds to friction between the intruder and the substrate. The digital readings are then converted to Newtons using empirically measured calibration rates. For this article, all force results are scaled by the friction force on the average particle, $\mu \langle m\rangle g$, where $\mu$ is the friction coefficient between the acetal rods and the acrylic substrate, $\langle m\rangle = (m_L + m_S)/2$ is the average mass of an individual rod, $m_L$ is the mass of a large rod, $m_S$ is the mass of a small rod, and $g$ is the acceleration due to gravity. To suppress noise in the force reading, we also apply Gaussian smoothing with a window size of $8$ frames.

\subsection{Image Analysis}
\label{sec:imageAnalysis}

In addition to measuring the global response of the granular system through force measurements, we also probe local structure response using image analysis and particle tracking. For both monomers and dimers, we track the tops of the circular rods, as well as the circular bumps on the intruder, using an edge detection algorithm~\cite{RieserThesis}. For this initial study, we simply analyze motion of the circular rods and do not track discrete dimer pairs as in Ref.~\cite{HarringtonPRE2018}. The position coordinates are also Gaussian smoothed, with a window size equal to the the number frames it takes for the intruder to move $\frac{2}{15}\frac{R}{v_p}$, where $v_p$ is the speed of the intruder and $R$ is the large rod radius. In addition to suppressing noise, we also use Gaussian smoothing to estimate the instantaneous velocity at each time frame.

We also use particle tracks to characterize local structure and deformation. For every time frame, we partition the granular packing and intruder `particles' into a radical Voronoi tessellation, implemented in {\tt voro++}~\cite{RycroftChaos2009}, and then a Delaunay triangulation. In Sec.~\ref{sec:SZ} and~\ref{sec:structure}, we describe how we use the Voronoi cells and triangles to characterize local dynamics and structure.

\section{Results}
\label{sec:results}

\subsection{Force Response}
\label{sec:forces}

\begin{figure}
\centering
\includegraphics[width=\linewidth]{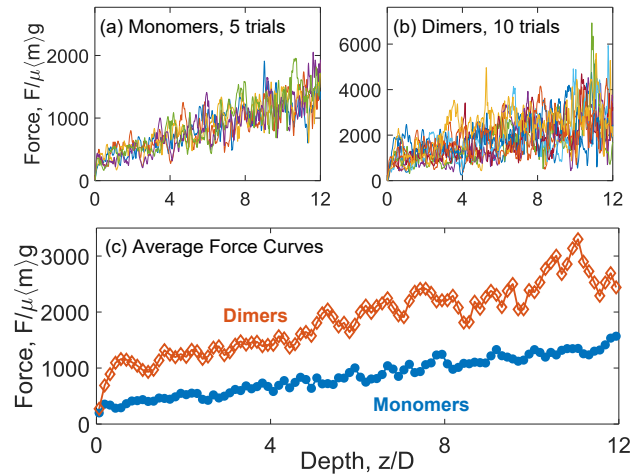}
\caption{Force response for individual trials of penetration into (a) monomers, (b) dimers. We perform 5 trials with monomer packings, 10 trials with dimer packings. (c) Average force response of monomers and dimers, measured over all trials taken for each shape. In all plots, force is scaled by $\mu\langle m\rangle g$, where $\mu$ is the friction coefficient between the grains and substrate, $\langle m\rangle = (m_L + m_S)/{2}$ is the average rod mass, $m_L$ is the large rod mass, $m_S$ is the small rod mass, and $g$ is the acceleration due to gravity. Depth is scaled by the large monomer diameter, $D$.}
\label{fig:forceResponse}
\end{figure}

We start with measurements of the force exerted on the intruder by the granular packing. Across individual trials, there is quite a bit of variation due to changes in the local structure while grains rearrange and avalanche. Fig.~\ref{fig:forceResponse}(a)-(b) shows force as a function of depth for individual trials of monomer and dimer packings. These plots alone would not suggest the presence of an overall force law.

However, a clear trend emerges when we average over all trials, binning force measurements by depth. Fig.~\ref{fig:forceResponse}(c) shows that both monomers and dimers exhibit a linear trend over a range of depths. While the curve for monomers remains consistently linear over the full range of depths, we observe deviation and noise in the dimer curve for depths beyond $4D$. As we describe in Sec.~\ref{sec:SZ}, we attribute this deviation to bottom boundary-influenced noise that has not been fully averaged out.

Despite the absence of hydrostatic pressure, we observe that the penetrative force increases linearly with depth. Under the guidelines of the local friction force model, the force needed to activate grains at a particular depth is proportional to the hydrostatic pressure, which varies linearly with depth. While this is absent in our apparatus, a similar local force picture arises from the sliding of grains that are above the current penetration depth. The number of grains above the intruder increases linearly, at the rate of the penetration speed, so this explains the presence of a linear force curve. 

Indeed, the two shapes appear to increase at similar rates, under equal penetration speeds. The primary differences between the shapes arises from the difference in yield, which is substantially larger for dimers. This suggests that while the long-term response for the shapes is similar, stagnant zone formation is highly affected by particle shape.

\subsection{Deviatoric Strain Rate \& Stagnant Zone}
\label{sec:SZ}

While we do not measure forces between grains in this apparatus, a close analogy to stress relaxation can be measured in the form of local strain or deformation. We quantify local strain over triangles obtained from Delaunay triangulation, including triangles that have constituent particles on the intruder. For a single triangle, we take the velocity, $v$, of each member at the vertices and obtain a local strain rate tensor $e$ using the constant strain triangle formalism~\cite{Cook1974},

\begin{equation}
	\left(\begin{array}{c}
		v_x(x,y)-v_{x,CM} \\
		v_y(x,y)-v_{y,CM}
	\end{array}\right)
	=
	\left(\begin{array}{cc}
		e_{11} & e_{12} \\
		e_{21} & e_{22}
	\end{array}\right)
	\left(\begin{array}{c}
		x \\
		y
	\end{array}\right),
\end{equation}
where $x$ and $y$ are Cartesian coordinates relative to the triangle centroid. From the empirical strain tensor $e$ we take its symmetric portion $\varepsilon$,

\begin{equation}
	\varepsilon_{ij} = \frac{e_{ij}+e_{ji}}{2}.
\end{equation}
Finally, we measure the deviatoric strain rate $J_2$,

\begin{equation}
\label{eq:J2}
	J_2 = \frac{1}{2}\sqrt{(\varepsilon_{11}-\varepsilon_{22})^2+4\varepsilon_{12}^2}.
\end{equation}

We use measurements of $J_2$ to quantify the degree to which local regions deform. As with measurements of max strain rate or velocity traces~\cite{AguilarNatPhys2016,MurthyPRE2012,ViswanathanGM2015}, $J_2$ serves as an indicator of the presence of a stagnant zone bounded by shear bands. We also use $J_2$ to assess the presence of boundary effects in our system, especially due to the bottom of the confining frame. We observe that, for both monomers and dimers, a significant $J_2$ signal does not emerge at the apparatus bottom until penetration depth exceeds $4D$. In Fig.~\ref{fig:forceResponse}(c), the average monomer force curve remains linear beyond this depth, while the dimer curve becomes increasingly noisy. Thus, to avoid boundary effects on local measurements, we restrict depths considered here and in Sec.~\ref{sec:structure} to within $D < z < 4D$, after the linear force law has set in, but before bottom boundary effects.

\begin{figure}
\centering
\includegraphics[width=\linewidth]{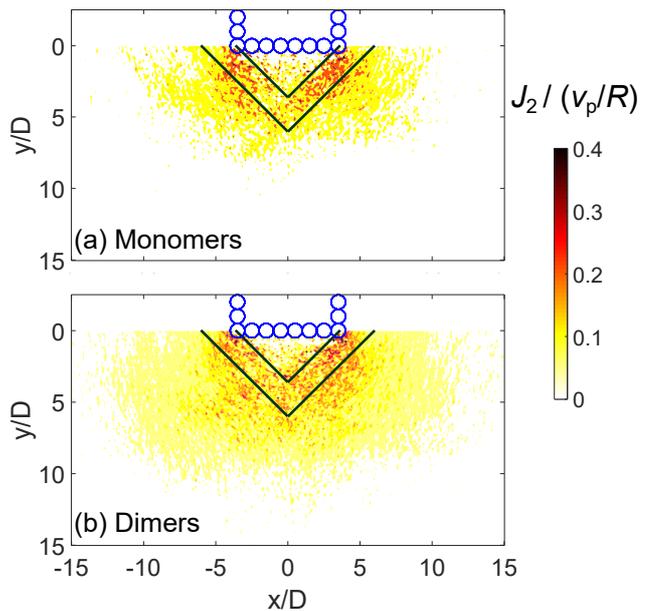}
\caption{Time-and trial-averaged spatial maps of $J_2$ in packings of (a) monomers, (b) dimers over the depth range $D < z < 4D$, restricted to an area in front of the intruder ($y/D < 15$, $|x|/D < 15$). We define Cartesian coordinates relative to the center of the bottom row of the intruder. The triangle bounded by the inner lines is the stagnant zone. The region between the inner and outer lines denotes locations of shear bands. These lines are drawn by hand to isolate the region of high, localized $J_2$ (shear band) and the region of low, sparse $J_2$ (stagnant zone).}
\label{fig:strain_SZ}
\end{figure}

Fig.~\ref{fig:strain_SZ} shows trial- and time-averaged maps of $J_2$ over Cartesian coordinates defined relative to the center of the bottom of the intruder. We define this reference point as the average position of intruder `particles' in the bottom row. For both shapes, there is a triangular stagnant zone marked by low and sparse measurements of $J_2$. Further from the intruder we see a backward-bending region of deformation, in line with characteristic paths drawn in Ref.~\cite{KangNatComms2018}. The extent of this region is substantially smaller for monomers, Fig.~\ref{fig:strain_SZ}(a), compared to dimers, Fig.~\ref{fig:strain_SZ}(b). Between these regions is a shear band interface with high $J_2$ that exhibits a similar shape as the stagnant zone.

We qualitatively distinguish these regions by drawing straight-line boundaries by hand. For both shapes, we use triangles that are centered at the middle of the intruder and have a fan angle of 90\degree ~($\frac{\pi}{2}$ rad). The theoretical model presented in Ref.~\cite{KangNatComms2018} relates this fan angle to the friction angle by means of a Mohr-Coulomb construction. A 90\degree ~fan angle in the stagnant zone would indicate grains with zero friction coefficient. Of course, there is actually a nonzero friction coefficient between the acetal surfaces of grains. However, the flat and highly frictional substrate changes the local failure criteria. Rather than requiring shear stress that overcomes frictional contacts and gravitational loading, the grains are primarily activated by overcoming static friction with the substrate. Not only does this simple penetration apparatus achieve a linear penetration force with a stagnant zone, but it also exhibits failure behavior that could serve as an experimental model for effectively frictionless particles.  

\subsection{Structure}
\label{sec:structure}

Finally, we measure structural anisotropies throughout the packing during penetration, and consider how they relate to local strain. As in the previous section, we limit our measurements to the depth range of $D < z < 4D$, after a stagnant zone has fully formed.

Structural anisotropies can be studied using a number of different approaches, from free volume~\cite{Turnbull1961}, to Voronoi cell size and shape~\cite{SchroderTurkEPL2010,MorseCorwinPRL2014}, to machine learning~\cite{CubukSchoenholzPRL2015,CubukIvancicScience2017}. For this study, we focus on a quantity that, like $J_2$, is defined over Delaunay triangles. Specifically, we measure the area-weighted divergence of the particle center-to-Voronoi cell centroid vector field, $Q_k$~\cite{RieserPRL2016}, 

\begin{equation}
\label{eq:Qk}
	Q_k = \nabla \cdot \mathbf{C}_k \frac{A_k}{\langle A \rangle}.
\end{equation}
By construction, the average $Q_k$ over an entire packing is zero. Positive values of $Q_k$ tend to correspond to closely packed, or ``overpacked,'' regions, while negative values correspond to void, or ``underpacked,'' regions.  Since the `intruder' particles do not have well defined Voronoi cells, we do not include triangles that contain them. We also note that for both monomers and dimers, as with $J_2$, we define $Q_k$ using triangles that consist of circular rods. In other words, we do not currently employ an approach that uses Voronoi cells surrounding bonded dimers.

We consider whether $Q_k$ illuminates distinct structural characteristics between the stagnant zone and shear band regions identified in Fig.~\ref{fig:strain_SZ}. This is motivated by the ability of $Q_k$ to distinguish structural behavior above and below jamming in simulated frictionless systems~\cite{RieserPRL2016}. Could a shear jammed region like the stagnant zone demonstrate similar structural signatures?

\begin{figure}
\centering
\includegraphics[width=0.80\linewidth]{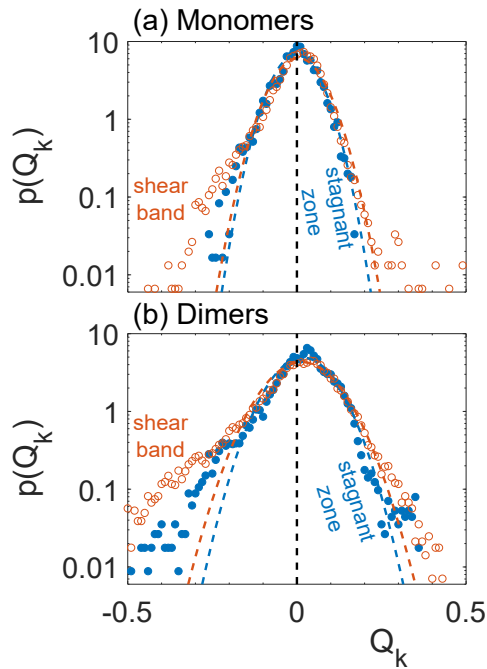}
\caption{Distributions of $Q_k$ measurements taken within the stagnant zone (closed circles) and shear band (open circles) regions for packings of (a) monomers, (b) dimers. The dashed lines are ideal Gaussian curves given the mean and variance measured over $-0.25 < Q_k < 0.25$. To mitigate boundary effects, we only consider measurements taken over the depth range from $D < z < 4D$.}
\label{fig:structure}
\end{figure}

In Fig.~\ref{fig:structure}, we show the distributions of $Q_k$ values found within the stagnant zone and the shear band. Indeed, for both shapes we observe distinct distributions of $Q_k$ between the two regions. Specifically, the stagnant zone has a more narrow distribution of $Q_k$ compared to the shear band. The width of a $Q_k$ distribution, quantified as the variance or standard deviation, is analogous to the inverse packing fraction. As expected, the deforming shear band is more dilated than the shear jammed stagnant zone. Moreover, the stagnant zone distributions exhibit different widths for monomers and dimers, indicating distinct packing fractions exhibiting shear jamming that are shape-dependent. 

\begin{figure}
\centering
\includegraphics[width=0.80\linewidth]{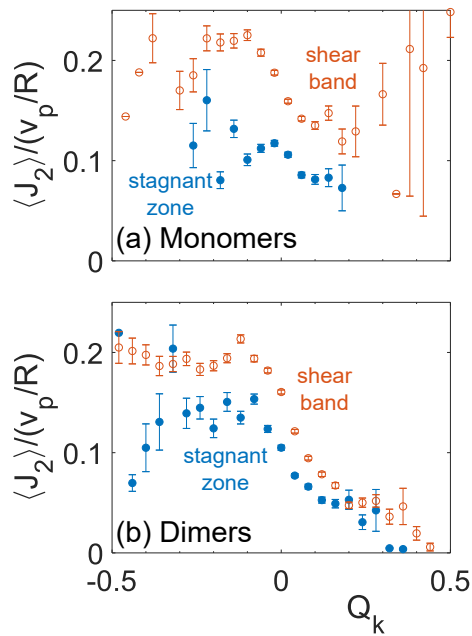}
\caption{Average $J_2$ measurements, as a function of corresponding $Q_k$, taken within the stagnant zone (closed circles) and shear band (open circles) regions for packings of (a) monomers, (b) dimers. Error bars represent the standard error of the mean $J_2$. To mitigate boundary effects, we only consider measurements taken over the depth range $D < z < 4D$.}
\label{fig:dynamics}
\end{figure}

Finally, we examine the relationship between structure and dynamics, as quantified by $Q_k$ and $J_2$. In general, an underpacked region ($Q_k < 0$) is expected to readily deform its open space, while an overpacked region ($Q_k > 0$) is more constrained and generally less likely to deform, up until the point where Reynolds dilatancy sets in. Fig.~\ref{fig:dynamics} shows the average relation between $Q_k$ and $J_2$ for the two grain shapes, in the shear band region and in the stagnant zone. Starting with monomers, we indeed observe this general trend within the shear band region, as well as a suppressed trend in the stagnant zone. Of course, this is in line with the stagnant zone exhibiting far fewer rearrangements and deformations. However, the dimers exhibit a similar relationship in \emph{both} regions. This highlights a distinct fragility in the stagnant zone for dimers, despite the observation that a similarly sized stagnant zone forms for both shapes. In other words, the structure of circular elements is less indicative of deformation.

\section{Discussion}
\label{sec:discussion}

In this article, we presented an experimental study of penetration into a 2D model granular bed made up of either circularly symmetric monomers or concave elongated dimers. The apparatus precludes hydrostatic pressure which is present in most granular penetration models. However, we still observe pertinent features of granular penetration. 

First, we find a global force response that is linear over a wide range of depths for the two shapes of grains. The rate of force growth is similar for both shapes, with dimers exhibiting a significantly stronger yield force. This signals that a linear depth-dependent force law is applicable in this system, despite the absence of hydrostatic pressure. Further study should elucidate better understanding of how dimers build up a larger yield force, including consideration of the effect of orientation and alignment~\cite{HarringtonPRE2018}. Moreover, individual runs exhibit avalanches that could be better understood relative to local rearrangements and structure.

Second, we directly observe a stagnant zone that forms in front of the intruding object, bounded by localized shear bands. The shape of the 2D stagnant zone is triangular, as expected, but with a fan angle that suggests that grain-grain interactions are effectively frictionless. We believe this may be due to the presence of a frictional substrate that is in flush contact with the flat rod bottoms. This acts as the primary activating interaction to overcome, rather than interparticle friction. To further explore other frictional effects, as well as other modes of particle interactions, particle rafts and highly frictional grains are worth investigating.

Finally, we directly probe local structure using a threshold-free metric that quantifies under- and overpacked regions. We find that the stagnant zone exhibits characteristics in the $Q_k$ distribution that are distinct from those in the surrounding shear band. We also find the particles shapes exhibit unique relationships between local structure and local strain within the stagnant zone. While deviatoric strain rate, $J_2$, is suppressed over all $Q_k$ values in the stagnant zone for monomers, dimers exhibit a clear negative correlation between the two quantities. This further illuminates distinct fragilities in the stagnant zones of the two shapes, and beckons further study into other aspects of structure. In particular, one can consider the structure of discrete dimer shapes, as well as the effects of elongated shapes that are convex, such as ellipses.

\begin{acknowledgements}
We thank Rafi Blumenfeld for helpful discussions, and Barbara Kountouzi and Kevin Paroda for fabrication assistance.  We also acknowledge Bob Behringer for his countless contributions to the field of granular matter, and to science as a whole.  This work was supported by NSF Grant MRSEC/DMR-1720530.
\end{acknowledgements}

\section*{Compliance with ethical standards}
\subsection*{Conflict of interest}
The authors declare that they have no conflict of interest.

\bibliographystyle{spphys}       
\bibliography{Penetration_bib}   

\end{document}